\documentstyle[12pt,preprint,aps,epsfig,floats,amssymb]{revtex}
\newcommand{\be}{\begin{equation}}
\newcommand{\ee}{\end{equation}}
\newcommand{\bea}{\begin{eqnarray}}
\newcommand{\eea}{\end{eqnarray}}
\newcommand{\ba}{\begin{array}}
\newcommand{\ea}{\end{array}}
\newcommand{\lsim}
{{\;\raise0.3ex\hbox{$<$\kern-0.75em\raise-1.1ex\hbox{$\sim$}}\;}}
\newcommand{\gsim}
{{\;\raise0.3ex\hbox{$>$\kern-0.75em\raise-1.1ex\hbox{$\sim$}}\;}}
\tighten
\begin{document}
\preprint{
\noindent
\begin{minipage}[t]{2in}
\begin{flushleft}
\end{flushleft}
\end{minipage}
\hfill
\begin{minipage}[t]{2in}
\begin{flushright}
IISc-CTS-18/01\\
\tt{hep-ph/0111273}\\
\vspace*{.2in}
\end{flushright}
\end{minipage}
}
\draft
\bibliographystyle{plain}
\thispagestyle{empty}
\vspace{-1cm}
\title{\bf  Reply to\\
Comment on Infrared Fixed Point Structure in the Minimal Supersymmetric
Standard Model 
with Baryon and Lepton Number Violation}

\bigskip

\author{B. Ananthanarayan}
\address{ Centre for Theoretical Studies, Indian Institute 
of Science, Bangalore 560 012, India}
\vspace{3mm}

\author{P. N. Pandita}
\address{ Department of Physics,
North-Eastern Hill University,  Shillong 793 022, India
}

\maketitle

\bigskip

\begin{abstract}
Infrared fixed points in the minimal supersymmetric
standard model with baryon and lepton number violation were studied 
and their structure elucidated by us in  
Physical Review {\bf D 63} 076008 (2001).  Here we reply to a comment
on this paper. We emphasize that our paper concentrates on the 
case of the only true infrared fixed point in the model, i.e. the stable
nontrivial fixed point for the 
top- and bottom-quark Yukawa couplings and the baryon number 
violating coupling. For this case the 
comment does not affect in any manner
the numerical results and conclusions derived in our paper.

\end{abstract}

\vspace{3mm}


\newpage

One of the main weaknesses of the standard model (SM), and its 
supersymmetric extensions, is that the masses of the 
matter particles, the quarks and leptons, are free parameters 
of the theory. This problem arises from the presence of many 
unknown dimensionless Yukawa couplings. Considerable attention has, therefore,
recently been focussed on the renormalization group (RG) evolution of the
various dimensionless Yukawa couplings in the SM 
and its minimal supersymmetric
extension, the minimal supersymmetric standard model (MSSM). Through the
RG evolution, one can relate the Yukawa couplings to the 
gauge couplings via the
Pendelton-Ross infrared stable fixed point (IRSFP) for the top-quark Yukawa
coupling, or via the quasi-fixed-point behavior~\cite{bs}. The 
predictive power of different models can, thus, be enhanced if the 
RG running of the parameters is dominated by the infrared stable fixed points.

\bigskip

In the minimal supersymmetric standard model, 
besides the Yukawa interactions  and their
supersymmetric counterparts, there are Yukawa couplings which 
violate baryon and lepton number conservation. 
This is in contrast to
the situation that one obtains in the SM, where gauge 
invariance, particle content and renormalizability forbid baryon and
lepton number violation at the level of renormalizable operators.
In~\cite{ba1}, following \cite{ba2}, 
we have carried out a detailed analysis of the infrared fixed point 
structure of the third generation Yukawa couplings, and the highest 
generation baryon and lepton number violating couplings in the 
minimal supersymmetric standard model. Having shown in~\cite{ba2} that
the only stable infrared fixed point of the model is the one with
nontrivial fixed point values for the top- and bottom-quark Yukawa couplings,
and the baryon number violating coupling $\lambda''_{233}$, and trivial 
fixed point values for rest of the  couplings, the purpose of~\cite{ba1} was
to obtain exact as well as approximate analytical solutions for these
Yukawa couplings and the baryon number violating coupling,
as well as the corresponding soft supersymmetry breaking
trlinear couplings at the weak scale given their values at the ultraviolet
scale. In~\cite{ba1}, 
we also studied the renormalization group 
flow of such a system, and
numerically determined the quasi-infrared-fixed surfaces and the 
quasi-infrared-fixed points
toward which the RG flow is attracted.

\bigskip

The main point of 
the authors of the comment~\cite{mm} is that the statement
``in the regime where the Yukawa couplings $\tilde Y_t(0), .....,
\tilde Y''(0) \rightarrow \infty $ with their
ratios fixed, 
it is legitimate to drop 1 in the denominators of Eq.(16) and
(18)--(23)......'' in~\cite{ba1} is not always correct.

\bigskip

As stated above, in our paper~\cite{ba1} we consider the 
quasi-infrared-fixed
points for the top- and bottom-quark Yukawa couplings, 
and the baryon number
violating coupling only, corresponding to the stable
nontrivial true fixed point for these couplings. 
In this case, we have
numerically obtained the quasi-fixed-point solution in our paper.
For the case of these couplings,
neglecting 1 in the denominator of
Eq.(16), and the associated Eqs.(18), (19), and (23),
of ~\cite{ba1}, in the quasi-fixed-point limit, is correct.
Furthermore, since we have not used Eq. (39), and Eqs. (40) - (45),
in~\cite{ba1} to obtain the  infrared point values for the case treated 
in our paper, but rather obtained them
directly by a numerical solution of the RG equations, 
the 1 is included in our study, i.e., the question of dropping 1 does 
not arise.
{\it In particular, the solutions for the relevant 
couplings around the true fixed point values, Eqs. (52) -- (61), and the 
quasi-fixed values for these couplings, (63) -- (64), 
obtained in our paper~\cite{ba1}
are correct.}

\bigskip

Although the case of all the couplings being large was 
mentioned in our paper~\cite{ba1}, we did  not study  the general case 
of quasi-infrared-fixed points where  all the Yukawa couplings  
go to infinity.   We concentrated  on  the case of top- and 
bottom-quark Yukawa couplings, and the baryon number violating coupling, 
since only these have been shown to have nontrivial 
true infrared fixed points~\cite{ba2}. For this case our analysis
is correct. For the rest of the couplings,
whose quasi-infrared-fixed-point behavior has not been studied in our
paper, a separate study might be needed as pointed out in~\cite{mm}.

\bigskip

In view of the above, the comment~\cite{mm}
is a technical statement about the validity
of Eq.(39) of our paper~\cite{ba1} for the general case 
when all the Yukawa
couplings become large. Although this equation was stated in our
paper, it was not used in the derivation of
our results, since our focus was on a different case.
As such, the comment~\cite{mm} does not alter the main results
and conclusions of our paper~\cite{ba1}.

\bigskip

{\bf Acknowledgements } 

\bigskip
 
The work of PNP is supported by  the University Grants Commission
under project No. 30-63/98/SA-III. He would like to thank 
the Abdus Salam International Center for Theoretical Physics, Trieste,
Italy for its hospitality while this work was completed.

\end{document}